\documentclass[showpacs,preprintnumbers,amsmath,amssymb,twocolumn,nobibnotes,longbibliography]{revtex4-2}
\usepackage[english]{babel}
\usepackage{mathtools}
\usepackage{graphicx}
\usepackage{mathrsfs}
\usepackage{amssymb}
\usepackage{xcolor}
\usepackage[colorlinks=true,citecolor=blue,urlcolor=blue]{hyperref}
\usepackage{amsmath}
\usepackage{bm}
\usepackage{physics}
\usepackage{slashbox}
\newcommand{\mysection}[1]{\textcolor{blue}{\textit{#1}.}}

\begin{document}

\title{Theory of weak localization in graphene with spin-orbit interaction}

\author{L. E. Golub}
\affiliation{Terahertz Center, University of Regensburg, 93040 Regensburg, Germany}

\begin{abstract}
Theory of weak localization in graphene with Rashba splitting of energy spectrum is developed. Anomalous magnetoresistance caused by weak localization is calculated with account for inter- and intravalley, spin-orbit and spin-valley scattering processes. It is shown that the anomalous magnetoresistance is described by the expression different from the traditional Hikami-Larkin-Nagaoka formula. The reason is that the effect of Rashba splitting gives rise to the spin-orbit vector potential which is not reduced to a spin dephasing only. The developed theory can be applied to  heterostructures of graphene with transition metal dichalcogenides.
\end{abstract}

\maketitle

\mysection{Introduction} 
Weak localization (WL) is an interference phenomenon based on wave properties of particles. 
WL mainly consists in interference of waves backscattered from groups of defects.
In conductors WL results in anomalous negative magnetoresistance in classically weak magnetic fields  at low temperatures. Study of magnetic-field and temperature dependence of the anomalous magnetoresistance allows determining various system parameters such as dephasing rates hardly accessible in other experiments~\cite{WL_review_Bergmann_1984}. In spin-orbit coupled systems, WL is much more reach due to constructive and destructive character of the interference in different spin channels described by additional spin-related phases acquired by electrons at backscattering. This results in sign-alternating magnetoresistance with a positive part at  lowest fields and named weak antilocalization (WAL). Studies of WAL in two-dimensional semiconductors give an access to spin-orbit splittings and spin relaxation times~\cite{WAL_SST_Review}.
Weak localization in graphene is specific due to the Berry phase $\pi$ acquired by Dirac fermions at backscattering~\cite{Graphene_review_2009}. However this does not always result in WAL due to effective intervalley scattering~\cite{McCann2006}.

In graphene-based systems the spin-orbit effects are important when the spin-orbit coupling is induced by proximity 
effects~\cite{SOI_graphene_Gmitra2015,Colloq_RMP2020}.
Theory for
graphene heterostructures predicts 
the Rashba~\cite{Rashba_1960}
spin-orbit splitting from 0.1 to a few meV in different graphene heterostructures with transition metal dichalcogenides (TMDC)~\cite{SOI_GR-TMDC_Fabian_2016,SR_anizotr_VZ-Rashba_Gr-TMDC_Fabian_2017,Gr-hBN_SOI_Fabian,SOI_GR-TMDC_twist_Koshino_2019} and topological insulators~\cite{SOI_twist_Gr-TI_heterostr_2023}.
Experiments on the anomalous magnetoresistance in graphene with spin-orbit coupling
 demonstrate  WAL in different single- and bilayer graphene/TMDC heterostructures~\cite{WL_Gr_TMDC_2015,WL_Gr_TMD_2016,WL_Gr_TMD_Eroms_2017,VZ_SOI_TMD-Gr-hBN_heterostr_exp_2018,WL_Gr_SOI_exp_2018,WL_Gr_SOI_exp_2019,WL_Gr_SOI_exp_2022,WAL_BLG_Amann2022}.
Determination of spin-orbit and dephasing parameters from experimental data is always performed by the theoretical expressions of Ref.~\cite{MF} containing the Hikami-Larkin-Nagaoka (HLN) function~\cite{HLN}.
However, it is known from studies of WAL in two-dimensional semiconductors with Rashba spin-orbit splitting that the HLN expression does not describe the anomalous magnetoresistance. Another formula derived by Iordanskii, Lyanda-Geller and Pikus should be used for description of experimental data~\cite{ILP,Knap}.
In this Letter we show that the same situation takes place in graphene with spin-orbit coupling and derive an expression for the WL induced anomalous magnetoresistance.

\mysection{Theory} 
The Hamiltonian of graphene with spin-orbit coupling
has the following form in the basis of eight-component Bloch functions of electrons on two lattices in two valleys and with one of two spin projections~\cite{McCann2006,MF,SR_aniz_2018}
\begin{multline}
\label{H}
\mathcal H = v \bm \Sigma \cdot \bm p +\gamma [\bm \Sigma \times \bm s]_z +\lambda \Sigma_z s_z
+\Delta_s\Pi_z\Sigma_z
\\ -\mu \Pi_z [\Sigma_x (p_x^2-p_y^2) - 2 \Sigma_y p_xp_y].
\end{multline}
Here $\bm p$ is momentum, $v$ is the Dirac fermion velocity, 
$z$ is a coordinate normal to the graphene layer,
the terms $\propto \gamma$ and $\propto \lambda$ are Rashba  and Kane-Mele (enhanced by phonons~\cite{Ochoa2012}) spin-orbit couplings, 
$\Delta_s$ is  an orbital gap due to staggered sublattice potential, and the term ${\propto \mu}$ describes trigonal warping.
We use three sets of Pauli matrices to describe spin $\bm s$,  sublattice ``isospin'' $\bm \Sigma$, and  matrices $\bm \Pi$ acting in the valley space~\cite{McCann2006}.
They are related to the Pauli matrices $\bm \sigma$ acting in the sublattice space by $\Sigma_{x,y}=\Pi_z \otimes \sigma_{x,y}$, 
$\Sigma_z=\sigma_z$.
Disorder is described by a sum of the spin-independent and spin-dependent terms~\cite{MF}
 \begin{multline}
V= uI + \sum_{a,l=x,y,z} u_{al}\Sigma_a \Lambda_l
\\
+\sum_{j=x,y,z} s_j \qty(\sum_{a=x,y,z}\alpha_{aj}\Sigma_a + \sum_{l=x,y,z}\beta_{lj}\Lambda_l),
\end{multline}
where valley ``pseudospin'' matrices are $\Lambda_{x,y}=\Pi_{x,y} \otimes \sigma_z$, $\Lambda_z=\Pi_z$.
The terms with $s_z$ and with $s_{x,y}$ describe the spin-orbit scattering due to $z \to -z$ symmetric and asymmetric perturbations, respectively~\cite{MF}.

In the low-temperature transport participate only the electrons at the Fermi level which is assumed to be far enough from the Dirac point. 
Therefore it is useful to pass to the new basis of electron states characterized by the momentum $\bm p$, which belong to a valley $K_\pm$.
If we consider the Hamiltonian $\mathcal H_0=v \bm \Sigma \cdot \bm p$ only, then we obtain that the energy of these 
states equals to $vp$ and the eigenfunction is $\ket{K_\pm,\bm p} = [1, \pm \exp(i\varphi_{\bm p})]^T/\sqrt{2}$,
where $\exp(i\varphi_{\bm p})=(p_x + ip_y)/p$.
In this new basis, the conduction-band Hamiltonian has the following form~\cite{Ilic2019}:
\begin{equation}
\label{H_c}
\mathcal H_c = vp + {\gamma \over p}[\bm p \times \bm s]_z
-\mu p^2 \cos{3\varphi_{\bm p}} \Pi_z,
\end{equation}
and matrix elements of  disorder scattering read
\begin{align}
\label{UV_matr_el}
&V_{\bm p' ,\bm p} = {\rm e}^{-i\theta/2} \biggl[ \cos({\theta/ 2}) \qty(u+ \sum_{l,j=x,y,z}\beta_{lj}\Pi_l s_j)
\\
&+\sum_{i,i'=x,y,z} \varkappa_i (\alpha_{ii'}s_{i'} + u_{ii'}\Pi_{i'}) + u{\varkappa_z \over \epsilon_{\rm F}} (\lambda s_z+\Delta_s \Pi_z)
\biggr]. \nonumber
\end{align}
Here $\epsilon_{\rm F}$ is the Fermi energy, $\theta = \varphi_{\bm p'} -\varphi_{\bm p}$ is the scattering angle, 
and we introduced a vector $\bm \varkappa(\bm p', \bm p) = [\cos{\Phi},-\sin{\Phi},i\sin({\theta/ 2})]$ where $\Phi=(\varphi_{\bm p'} +\varphi_{\bm p})/2$.
The Kane-Mele term 
$\propto \lambda$ and the staggered-potential term $\propto \Delta_s$ in Eq.~\eqref{H}
mix the  conduction- and valence-band states in each valley which results in  spin- and valley-dependent scattering corrections.

It follows from Eqs.~\eqref{H_c},~\eqref{UV_matr_el} that  the Hamiltonian in both valleys has the same form as
that for spin-splitted massive electrons 
with angle-dependent scattering, the only difference is the factor ${\rm e}^{-i\theta/2}$ in the scattering amplitude.
Therefore the problem of WL in graphene with spin-orbit splitting is reduced to the problem of WL 
of massive electrons with Rashba splitting
solved in Ref.~\cite{ILP}.
For 
the Hamiltonian~\eqref{H_c}, equation for the Cooperon $\mathcal C(\bm q)$ 
has the form~\cite{SM}
\begin{multline}
\label{Cooperon_eq}
\biggl\{Dq^2+ \Gamma_\phi+\Gamma +\Gamma_{\rm R} \qty(\bm S^2- S_z^2) 
\\
+\sqrt{2 \Gamma_{\rm R}\tau_{\rm tr}} [\bm S \times \bm q]_z v
\biggr\}\mathcal C(\bm q)=1.
\end{multline}
Here $\bm q$
is the generalized momentum of a double charge in magnetic field, 
$\Gamma_\phi$ is the spin- and valley-independent dephasing rate,
$D=v^2\tau_{\rm tr}/2$ is the diffusion coefficient,
where $\tau_{\rm tr}^{-1}=\pi g u^2/(2\hbar)$ is the transport relaxation rate and $g=\epsilon_{\rm F}/(2\pi\hbar^2 v^2)$ is the density of states at the Fermi energy per spin per valley,
 $\Gamma$ is the dephasing operator including the effect of warping from the last term in Eq.~\eqref{H_c} but independent of the Rashba splitting,
$\Gamma_{\rm R}=2(\gamma/\hbar)^2\tau_{\rm tr}$ is the Rashba-term induced Dyakonov-Perel spin relaxation rate, and 
$\bm S$ is the operator of total spin of two interferring particles.
It is crucial that 
the bilinear in momentum and spin 
term $\propto [\bm S \times \bm q]_z$ 
is present in the  Cooperon equation which 
mixes different spin interference contributions to conductivity.
The $\Gamma_{\rm R}$-related terms in Eq.~\eqref{Cooperon_eq} mean that the spin-orbit splitting results in a spin-orbit vector potential ${\bm A_{\rm SO} = (\gamma/ev) [\hat{\bm z}\times \bm S }]$ which is not reduced to spin dephasing only~\cite{SM}.

\begin{table*}[t]
	\caption{Dephasing rates $\Gamma^{l}_{j}$ for different valley and spin channels with $l,j=s,t_0,t_1$.
Here 
$\Gamma_{\rm KM}= {2\pi\over \hbar}g \sum_{l=x,y,z} \alpha_{lz}^2 + {2\pi\over \hbar}g\qty(u{\lambda/ \epsilon_{\rm F}})^2$, $\Gamma_\alpha = {2\pi\over \hbar}g \sum_{l=x,y,z}\alpha_{lx}^2$,
$\Gamma_z = {2\pi\over \hbar}g \sum_{l=x,y,z}u_{lz}^2 + \Gamma_w+ {2\pi\over \hbar}g\qty(u{\Delta_s/ \epsilon_{\rm F}})^2$,
where $\Gamma_w=(\mu \epsilon_{\rm F}^2/\hbar v^2)^2\tau_{\rm tr}$,
${\Gamma_{iv} = {2\pi\over \hbar}g \sum_{l=x,y,z}u_{lx}^2}$,
$\{ \Gamma_{zv,e}, \Gamma_{iv,e}, \Gamma_{zv,o}, \Gamma_{iv,o}\} = {\pi g\over \hbar} \{\beta_{zz}^2, \beta_{xz}^2, \beta_{zx}^2, \beta_{xx}^2\}$,
$\Gamma_*=\Gamma_z+\Gamma_{iv}$  and
$\Gamma_{\rm SO}=\Gamma_{\rm asy} +\Gamma_{\rm sym}$,
where
$\Gamma_{\rm asy}=\Gamma_\alpha+2\Gamma_{zv,o}+ 4\Gamma_{iv,o}$
and
$\Gamma_{\rm sym}=\Gamma_{\rm KM} +2\Gamma_{zv,e} + 4\Gamma_{iv,e}$.
}
	\label{tab_rates}
\begin{tabular}{|c|c|c|c|}
\hline
\hline
\backslashbox{spin}{valley}
      & $t_1$ & $t_0$ & $s$ \\ \hline
$t_1$ &  $\Gamma_* +  \Gamma_{\rm KM} + 2\Gamma_{iv,e}  + \Gamma_\alpha + 2\Gamma_{zv,o} + 4\Gamma_{iv,o}$ 
& $2\Gamma_{iv}+  \Gamma_{\rm KM} +  2\Gamma_{zv,e} + \Gamma_\alpha+ 2\Gamma_{zv,o} + 4\Gamma_{iv,o}$      
& $\Gamma_{\rm SO}$   \\
$t_0$ &  $\Gamma_* +  2(\Gamma_\alpha + \Gamma_{zv,e} + \Gamma_{iv,e} +2\Gamma_{iv,o})$     
& $2(\Gamma_{iv}+\Gamma_\alpha + 2\Gamma_{zv,o}+2\Gamma_{iv,e})$    
& $2\Gamma_{\rm asy}$    \\
$s$    &  $\Gamma_* +  2(\Gamma_{zv,e} +2\Gamma_{zv,o} + \Gamma_{iv,e} +2\Gamma_{iv,o})$     
& $2(\Gamma_{iv}+2\Gamma_{iv,e}+\Gamma_{iv,o})$         &0   
\\
\hline
\hline
\end{tabular}
\end{table*}

There are valley and spin singlet ($s$) and triplet ($t_0$, $t_1$) interference channels of WL where
$t_0$ and $t_1$ correspond to spin/pseudospin $z$-projection equal to zero or $\pm 1$, respectively~\cite{zero_field_2022}.
Projecting the operator $\Gamma$ in Eq.~\eqref{Cooperon_eq} onto these
states~\cite{SM}, we obtain,
in addition to 
$\Gamma_\phi$, nine dephasing rates 
$\Gamma^{l}_{j}$ for valley $l$ and spin $j$
channels with $l,j=s,t_0,t_1$
given in the Table~\ref{tab_rates}~\footnote{In the notations of Ref.~\cite{MF}, $\Gamma^{l}_{j}=\Gamma^m_n$ where $l,j=s,t_0,t_1$ correspond to $m,n=0,z,x$.}.
Note that, by contrast with
  Ref.~\cite{MF}, $\Gamma_{\rm R}$ is not added to $\Gamma_\alpha$ because it
  enters into the Cooperon equation~\eqref{Cooperon_eq} 
not only as quadratic but also as linear in $\bm S$ terms,
  and is an independent parameter of the theory.

Solving Eq.~\eqref{Cooperon_eq} in magnetic field $\bm B$ normal to the graphene layer 
we obtain the WL induced magnetoconductivity $\Delta\sigma=\sigma(B)-\sigma(0)$
in the following form~\cite{SM}
\begin{multline}
\label{sigma_spin+valley+spin-valley}
{\Delta\sigma \over \sigma_0} 
= -\sum_{l=t_1,t_0,s} c_l \biggl[ \mathcal F_t\qty({\mathcal B_\phi\over B},{\mathcal B_{\rm R}\over B},{\mathcal B_{t_1}^{l}\over B},{\mathcal B_{t_0}^{l}\over B}) 
\\ - F\qty({B\over \mathcal B_\phi + \mathcal B_{s}^{l}}) \biggr].
\end{multline}
Here $\sigma_0=e^2/(2\pi h)$,  
 the common negative sign is caused by the Berry phase $\pi$ of Dirac fermions, 
 $\{ \mathcal B_{{\rm R}}, \mathcal B_\phi, \mathcal B_{j}^{l}\}=\{ \Gamma_{{\rm R}}, \Gamma_\phi, \Gamma_{j}^{l}\}\hbar/(4\abs{e}D)$,
and 
$c_{t_1}=2$, $c_{t_0}=1$, $c_s=-1$.
The singlet contribution is expressed via  the HLN function $F(x)=\psi(1/2+1/x)+\ln{x}$ with the digamma function $\psi(y)$.
By contrast, the triplet contribution is given by a four-parametric function
\begin{widetext}
\begin{equation}
\label{F_t}
\mathcal F_t\qty(b_\phi,b_{\rm R},b_1,b_0) = 
\sum_{m=0,\pm}  \qty[u_m \psi\qty(1/2 + b_\phi  + \bar{b} - v_m)  - u_m^{(0)}\ln{\qty(b_\phi  + \bar{b} - v_m^{(0)})}]
+{1\over (b_\phi+b_{\rm R} + b_{1})^2-1/4}.
\end{equation}
\end{widetext}
Here 
\begin{equation}
\bar{b} = (b_{0}+2b_{1})/ 3, \quad b_-=(b_{0}-b_{1})/3,
\end{equation}
and
the coefficients $u_m$, $v_m$ are found by the method of Punnoose~\cite{Punnoose,Punnoose-Manfra}:
\begin{align}
\label{vs_us}
&v_m = 2\delta \cos(\varphi + {2\pi\over 3}m), \quad \delta= \sqrt{-{C/ 3}}, 
\\ 
&\varphi = {1\over 3}\arccos({-{G\over \delta^3}})-{2\pi\over 3}, \nonumber
\quad u_m = {3v_m^2 + 4b_{\rm R}v_m +A\over \prod_{m'\neq m}(v_m-v_{m'})},\nonumber
\end{align}
where
\begin{multline}
A=5b_{\rm R}^2+4b_{\rm R}(b_\phi+\bar{b})-1  - b_-(3b_-+2b_{\rm R}),
\\
C=A-4b_{\rm R}(b_{\rm R}-b_-), \quad {G}=2b_{\rm R}(b_\phi+\bar{b})(b_{\rm R}-b_-) 
\\
+ b_{\rm R}^3 - b_-( b_{\rm R}b_-   -b_-^2+b_{\rm R}^2+1).
\end{multline}
The coefficients $v_m^{(0)}$ and $u_m^{(0)}$ are calculated by Eqs.~\eqref{vs_us} with the zero-field asymptotes of $A$, $C$ and $G$ given by $A^{(0)}=A+1$, $C^{(0)}=C+1$, $G^{(0)}=G+b_-$.

\mysection{Discussion} 
In the absence of Rashba splitting, $\mathcal B_{\rm R}=0$, we have 
$A=C=-1-3b_-^2$ and ${G=b_-(b_-^2-1)}$. This yields
$u_{0,\pm}=1$, $\{v_{0},v_+,v_-\}= { \{b_- -1, b_- +1, -2b_-\} }$, hence 
the triplet contribution is simplified to 
$\mathcal F_t \qty(b_\phi,0,b_1,b_0) = 
2F\qty(b_\phi+b_1)+F\qty(b_\phi+b_0)$,
and we obtain the HLN-like expression
\begin{multline}
{\Delta\sigma \over \sigma_0} \biggr|_{\mathcal B_{\rm R}=0}= 
-\sum_{l=t_1,t_0,s} c_l \biggl[2F\qty({B\over \mathcal B_\phi+\mathcal B_{t_1}^{l}})
\\ +F\qty({B\over \mathcal B_\phi+\mathcal B_{t_0}^{l}})  - F\qty({B\over \mathcal B_\phi + \mathcal B_{s}^{l}})\biggr].
\end{multline}

\begin{figure}[h]
	\centering \includegraphics[width=0.75\linewidth]{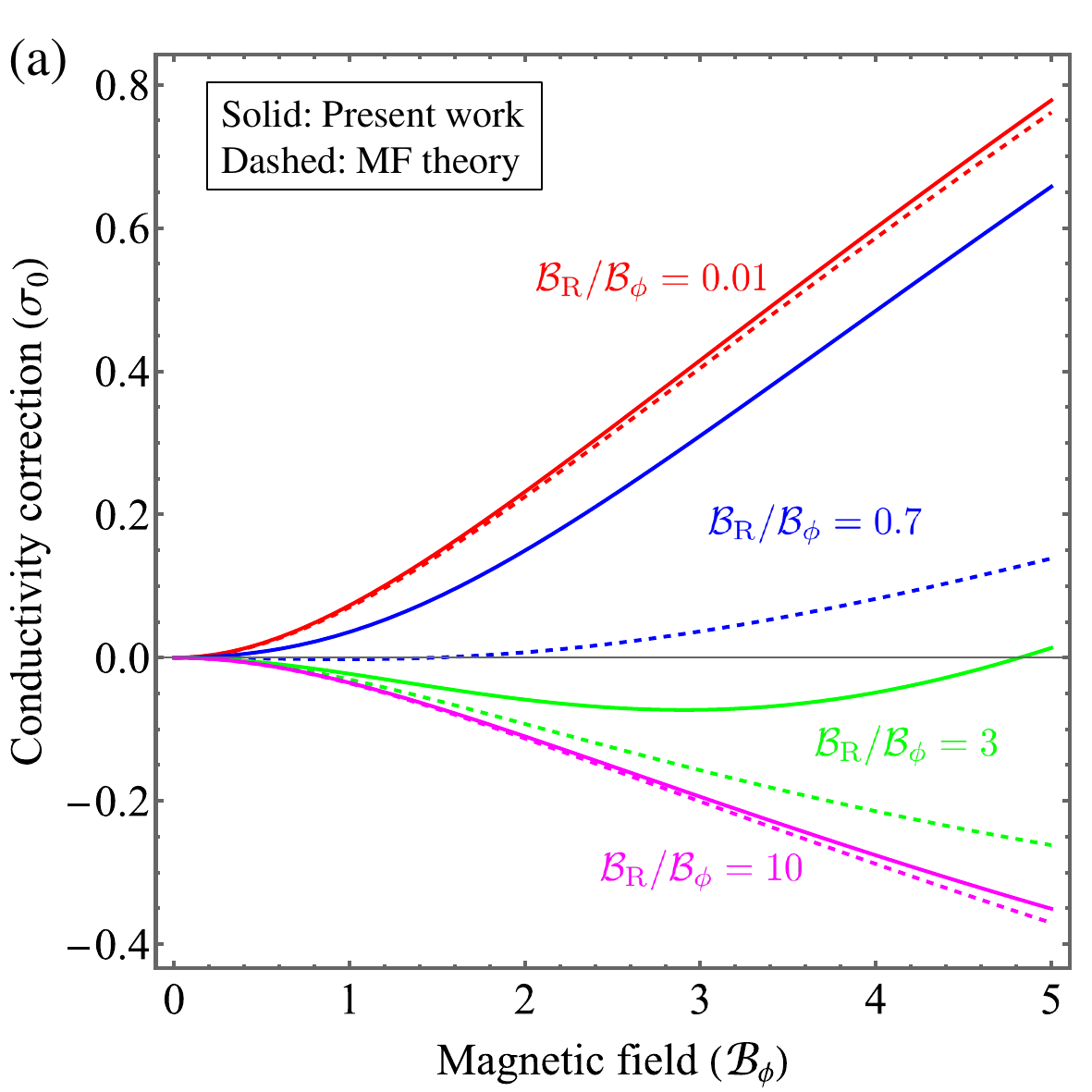} 
	\centering \includegraphics[width=0.75\linewidth]{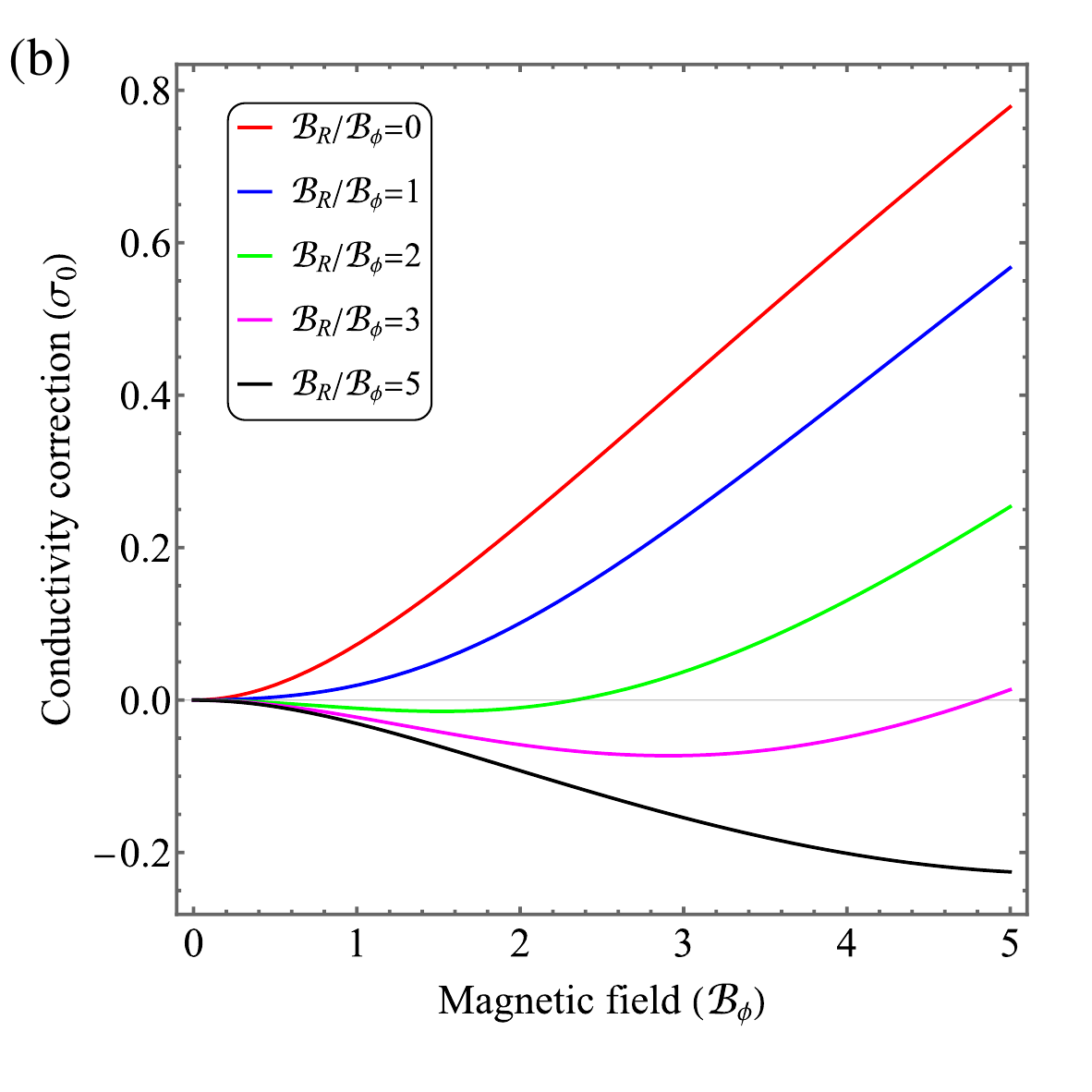}
	\caption{ 
	Conductivity correction 
	in the absence of spin-orbit scattering.
(a): present work Eq.~\eqref{valley-singlet_only} (solid lines) and MF theory Eq.~\eqref{MF} (dashed).
	(b): 
	Eq.~\eqref{valley-singlet_only} 
	at different values of $\mathcal B_{\rm R}/\mathcal B_\phi$.}
	\label{Fig_ILP-HLN}
\end{figure}

Let us 
consider a limit of fast valley-triplet relaxation: $\Gamma_{j}^{t_{1,0}} \gg \Gamma_{j}^{s}$. Then the anomalous magnetoresistance is described by just three dephasing rates, $\Gamma_\phi$, $\Gamma_{\rm SO}$ and $\Gamma_{\rm asy}$:
\begin{equation}
\label{valley-singlet_only}
{\Delta\sigma \over \sigma_0} \biggr|_{\Gamma_{j}^{t_{1,0}} \gg \Gamma_{j}^{s}}= \mathcal F_t\qty({\mathcal B_\phi\over B},{\mathcal B_{\rm R}\over B},{\mathcal B_{\rm SO}\over B},{\mathcal B_{\rm asy}\over B})- F\qty({B \over \mathcal B_\phi}),
\end{equation}
where $\mathcal B_{\rm SO}=\Gamma_{\rm SO}\hbar/(4\abs{e}D)$ and $\mathcal B_{\rm asy}=\Gamma_{\rm asy}\hbar/(2\abs{e}D)$.
If, in addition, $\mathcal B_{\rm R}=0$ then,
combining two previous equations,
we obtain the expression of Ref.~\cite{MF}
\begin{multline}
\label{zero_Rashba}
{\Delta\sigma(B) \over \sigma_0}\biggr|_{\mathcal B_{\rm R}=0,\Gamma_{j}^{t_{1,0}} \gg \Gamma_{j}^{s}} = 2F\qty({B\over \mathcal B_\phi +\mathcal B_{\rm SO}})
\\ +F\qty({B\over \mathcal B_\phi +\mathcal B_{\rm asy}})- F\qty({B\over \mathcal B_\phi }).
\end{multline}

McCann and Fal'ko (MF) derived in Ref.~\cite{MF} an expression, which, 
 in the absence of spin-orbit scattering
 but in the presence of Rashba splitting, gives:
\begin{equation}
\label{MF}
{\Delta\sigma_{\rm MF} \over \sigma_0}
= 2F\qty({B\over \mathcal B_\phi +\mathcal B_{\rm R}})
+F\qty({B\over \mathcal B_\phi +2\mathcal B_{\rm R}})- F\qty({B\over \mathcal B_\phi }).
\end{equation}
This expression treats the Rashba spin-orbit coupling as dephasing only and ignores the effect of the linear in $\bm S$ terms in the Cooperon equation.
Theory developed in the present work gives for this case Eq.~\eqref{valley-singlet_only} where $\mathcal B_{\rm SO}=\mathcal B_{\rm asy}=0$. 
In Fig.~\ref{Fig_ILP-HLN}~(a) we compare these two expressions 
demonstrating that they differ significantly.
The MF expression is correct either in the absence of spin-orbit coupling ($\mathcal B_{\rm R}=0$) or when it is very large and the triplet contribution is negligible ($\mathcal B_{\rm R} \gg \mathcal B_\phi$).
At intermediate values of the ratio $\mathcal B_{\rm R} / \mathcal B_\phi$, the expression derived in the present work 
shows that strong WAL effect is already present at moderate Rashba splitting when $\mathcal B_{\rm R} \geq 3\mathcal B_\phi$.
In Fig.~\ref{Fig_ILP-HLN}~(b) we show the magnetoconductivity in the absence of spin-orbit scattering 
at different values of the Rashba splitting.

\begin{figure}[h]
	\centering \includegraphics[width=0.75\linewidth]{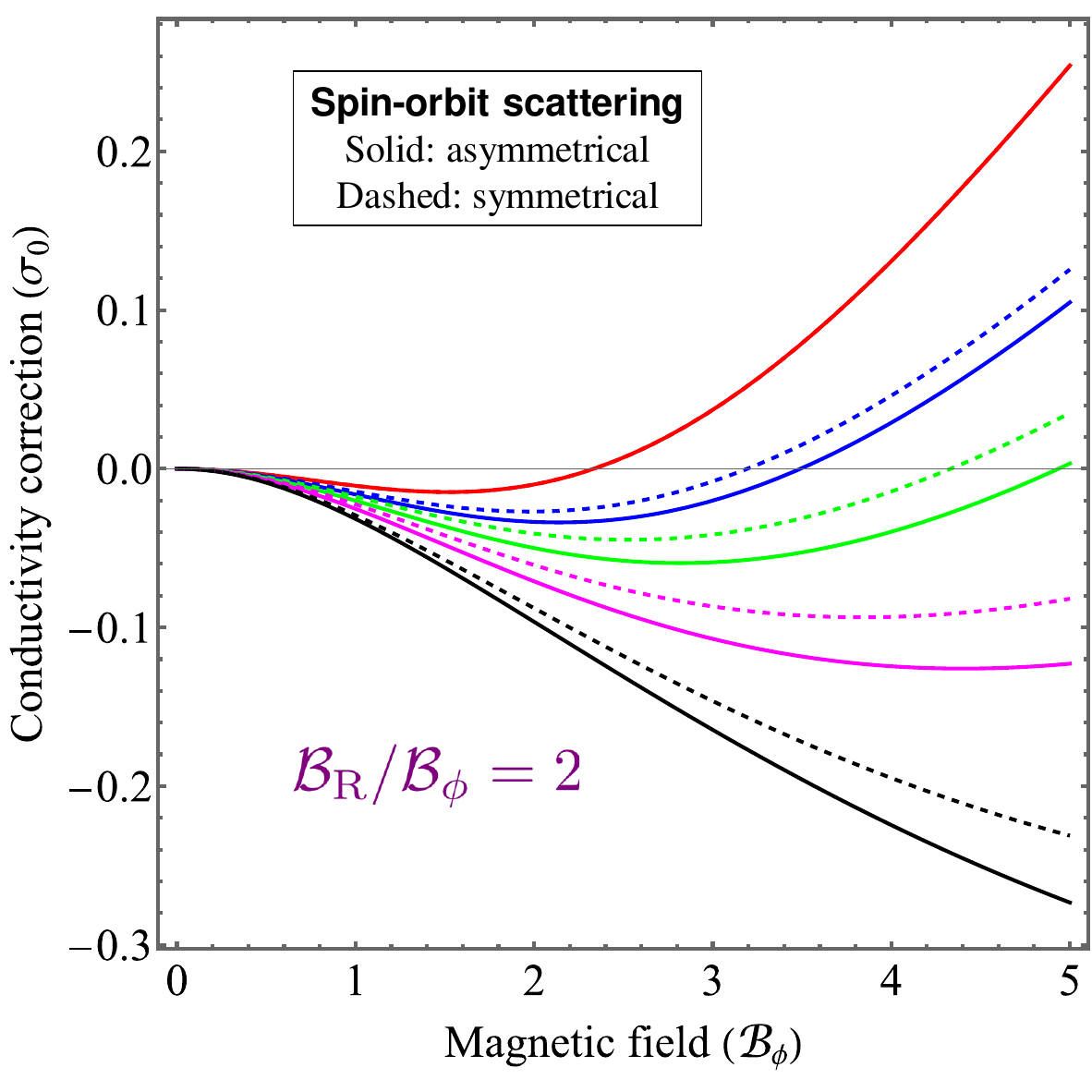} 
	\caption{ 
	Effect of spin-orbit scattering at $\mathcal B_{\rm R}=2\mathcal B_\phi$.
Solid (dashed) lines: 
$\mathcal B_{\rm asy}(\mathcal B_{\rm sym}) =0.5\mathcal B_\phi$~(blue), 
$\mathcal B_\phi$~(green), $2\mathcal B_\phi$~(magenta) and $5\mathcal B_\phi$~(black).
Red: $\mathcal B_{\rm asy} = \mathcal B_{\rm sym}=0$.
}
	\label{Fig_SO_scatt_effect}
\end{figure}

Effect of spin-orbit scattering is demonstrated in Fig.~\ref{Fig_SO_scatt_effect}. In the presence of a moderate Rashba splitting ${\mathcal B_{\rm R}=2\mathcal B_\phi}$, both symmetrical and asymmetrical spin-orbit scattering processes result in a more anti-localizing behaviour of the magnetoresistance. 
We considered asymmetrical ($\mathcal B_{\rm sym}=0$, $\mathcal B_{\rm SO}=\mathcal B_{\rm asy}/2$) and symmetrical  ($\mathcal B_{\rm asy}=0$, $\mathcal B_{\rm SO}=\mathcal B_{\rm sym}$) scattering.
Figure~\ref{Fig_SO_scatt_effect} shows that the effect of the asymmetrical scattering is stronger.

In an in-plane magnetic field, the Zeeman term 
${(\epsilon_{\rm Z}/2)\bm s \cdot \bm l_\parallel}$ is added to the Hamiltonian, where $\epsilon_{\rm Z}$ is the Zeeman splitting and $\bm l_\parallel$ is a unit vector in the direction of the field. 
It results in a mixing of singlet and triplet spin Cooperons in addition to the pure triplet mixing by the spin-orbit vector potential~\cite{Malshukov1997,wl_tilted}. 
WL conductivity correction calculation
in the presence of both  Rashba and in-plane Zeeman splittings~\cite{WAL_SST_Review} 
showed
that the conductivity as a function of in-plane magnetic field 
has a maximum when Zeeman and Rashba splittings are equal, $\epsilon_{\rm Z} \approx 2\gamma$.
In moderate fields where the Zeeman splitting is smaller than the spin relaxation gaps, $\epsilon_{\rm Z} \ll \hbar \Gamma_{\rm R, SO}$, the effect of the parallel field mainly consists in a suppression of the spin singlet interference channel~\cite{Malshukov1997,wl_tilted}. For graphene with spin-orbit interaction, this Zeeman splitting induced dephasing rate is $\Gamma_s^s=(\epsilon_{\rm Z}/\hbar)^2/(\Gamma_{\rm R}+\Gamma_{\rm SO})$, and  Eq.~\eqref{valley-singlet_only} yields
\begin{equation}
\label{dsigma_with_Gamma_ss}
{\Delta\sigma \over \sigma_0}= \mathcal F_t\qty({\mathcal B_\phi\over B},{\mathcal B_{\rm R}\over B},{\mathcal B_{\rm SO}\over B},{\mathcal B_{\rm asy}\over B})- F\qty({B \over \mathcal B_\phi + \mathcal B_s^s}),
\end{equation}
where we again assumed valley-triplet channels to be suppressed, $\Gamma_{j}^{t_{1,0}} \gg \Gamma_{j}^{s}$.
In the opposite limit of large Zeeman splitting $\epsilon_{\rm Z} \gg \gamma$, Dyakonov-Perel spin relaxation is suppressed, and the magnetoconductivity is given by the expression with $\Gamma_{\rm R} \to 0$, Eq.~\eqref{zero_Rashba}.

The valley-Zeeman term $\lambda_{\rm VZ} \Pi_z s_z$~\cite{VZ_SOI_TMD-Gr-hBN_heterostr_exp_2018,SR_aniz_2018} makes a similar effect if intervalley scattering is ineffective. The spin-singlet dephasing rate $\Gamma_s \sim (2\lambda_{\rm VZ}/\hbar)^2/(\Gamma_{\rm R}+\Gamma_{\rm SO})$ appears for each valley resulting in the magnetoconductivity given by Eq.~\eqref{dsigma_with_Gamma_ss} multiplied by the factor $-2$.
In the opposite limit of  comparable inter- and intravalley scattering efficiencies the valley-Zeeman splitting has no effect on WL.

The pseudospin inversion asymmetry terms in the Hamiltonian bilinear in spin and momentum~\cite{SOI_GR-TMDC_Fabian_2016,Gr-hBN_SOI_Fabian} result in small renormalizations of the Rashba constant and the spin relaxation rate $\Gamma_\alpha$.

The theory presented above gives WL magnetoconductivity in magnetic field smaller than the ``transport'' field $\mathcal B_{\rm tr}=\hbar/(4eD\tau_{\rm tr})$. In higher fields $B \sim \mathcal B_{\rm tr} \gg \mathcal B_\phi$, interference on ballistic trajectories with a few number of scatterers is important in WL. In this case,
the non-diffusive theory can be developed accounting for both the Dirac fermion nature of carriers~\cite{GKO,MN_NS} and the spin-orbit coupling~\cite{WAL_SST_Review} as well as valley-Zeeman splitting of arbitrary strengths.

In bilayer graphene with spin-orbit coupling, the spin-orbit vector potential in Eq.~\eqref{Cooperon_eq} is quadratic in $q$. Therefore it has no significant effect on the anomalous magnetoresistance, and the HLN-like theory~\cite{WAL_BLG_Amann2022} is correct.
The same is true for WL in TMDC layers where the spin-orbit vector potential is absent due to lack of linear in momentum terms in the Hamiltonian~\cite{WL_TMDC}.

\mysection{Conclusion} 
The developed theory of WL in graphene accounts for the Rashba spin-orbit splitting, spin-orbit and valley-dependent scattering. It is shown that the Rashba interaction affects WL in graphene not via spin dephasing but via a spin-orbit vector potential.
This results in the expression for the anomalous magnetoconductivity which is not reduced to the traditional formulas with HLN functions. An importance of this difference is demonstrated. The present theory allows one to determine  adequately the spin-orbit parameters of graphene with spin-orbit interaction from experimental data.

This work was supported by the Deutsche Forschungsgemeinschaft (DFG, German Research Foundation) via Project-ID 314695032 -- SFB 1277 (Subproject  A04) and the Volkswagen Stiftung Program (97738).

\bibliography{WAL_Gr_SOI}

\onecolumngrid

\newpage

\begin{center}
\makeatletter
{\large\bf{Supplemental Material for\\``\@title''}}
\makeatother
\end{center}

\let\oldsec\section

\renewcommand{\thesection}{S\arabic{section}}
\renewcommand{\section}[1]{\oldsec{#1}}
\renewcommand{\thepage}{S\arabic{page}}
\renewcommand{\theequation}{S\arabic{equation}}
\renewcommand{\thefigure}{S\arabic{figure}}

\setcounter{page}{1}
\setcounter{section}{0}
\setcounter{equation}{0}
\setcounter{figure}{0}

\section{Cooperon equation}

\label{App_Cooperon}

The Cooperon in the presence of momentum-dependent energy splittings as well as spin- and valley-dependent scattering is found from the following equation~\cite{ILP,Knap,WL_ftp}
\begin{multline}
C(\bm p', \bm p, \bm q) = {2\pi g \tau_0^2\over \hbar} V_{\bm p', \bm p} \otimes V_{-\bm p', -\bm p} 
\\+ {2\pi g \over \hbar} \expval{V_{\bm p', \bm k} \otimes V_{-\bm p', -\bm k}  C(\bm k, \bm p, \bm q) \qty(\tau_0^{-1} - i v\bm n_{\bm k}\cdot \bm q - i  {2\gamma\over \hbar}[\bm n_{\bm k} \times \bm S]_z +i{2\mu k^2\over \hbar} \cos{3\varphi_{\bm k}}J_z - \Gamma_\phi)^{-1}}_{\varphi_{\bm k}}.
\end{multline}
Here $\bm n_{\bm k}=\bm k/k$, $\bm S=(\bm s_1 + \bm s_2)/2$ and $\bm J=(\bm \Pi_1 + \bm \Pi_2)/2$ are operators of total spin and pseudospin of two interferring particles, angular brackets denote averaging over directions at the Fermi circle, and $\tau_0^{-1}=(\pi g/2\hbar){\rm Tr}\expval{V_{\bm p', \bm p}^\dag V_{\bm p', \bm p}}_\theta$, where $\rm Tr$ denotes a trace over both spin and valley indices.
The WL correction to the conductivity is expressed via the Cooperon as follows
\begin{equation}
\label{SM_cond}
\sigma=-2\pi D \sigma_0 \sum_{\bm q} \sum_{l,j=t_1,t_0,s}c_l c_j\expval{C^l_j(-\bm p, \bm p, \bm q)}_{\varphi_{\bm p}},
\end{equation}
where 
$C^l_j$ is a projection of the Cooperon on the $l$th valley and $j$th spin interference channel, and $c_{t_1}=2$, $c_{t_0}=1$, $c_s=-1$.

In graphene with spin-orbit interaction, the scattering amplitude [Eq.~(4) of the main text] has the form
\begin{equation}
V_{\bm p', \bm p} = {\rm e}^{-i\theta/2} \qty[u \cos({\theta/ 2}) +U_{\bm p', \bm p}],
\end{equation}
where $\theta = \varphi_{\bm p'}-\varphi_{\bm p}$, and a small correction $U_{\bm p', \bm p}$ is due to spin-orbit, valley-orbit and spin-valley scattering.
Therefore the Cooperon equation reads
\begin{multline}
C(\bm p', \bm p, \bm q) = 2\tau_0\cos^2\qty({\theta/ 2}){\rm e}^{-i\theta} 
+ {2\pi g \tau_0\over \hbar}{\rm e}^{-i\theta}\expval{U_{\bm p', \bm k} \otimes U_{-\bm p', -\bm k} {\rm e}^{i\theta'} C(\bm k, \bm p, \bm q) }_{\varphi_{\bm k}}
\\+ {2\pi g\over \hbar}{\rm e}^{-i\theta} \expval{u^2 \cos^2\qty({\theta-\theta'\over 2}) {\rm e}^{i\theta'} C(\bm k, \bm p, \bm q) \qty{\tau_0^{-1} 
- i \bm n_{\bm k}\cdot \qty( v\bm q + {2\gamma\over \hbar}[\bm S \times \hat{\bm z}]) 
+i{2\mu k^2\over \hbar} \cos{3\varphi_{\bm k}}J_z - \Gamma_\phi }^{-1}}_{\varphi_{\bm k}},
\end{multline}
where $\theta'=\varphi_{\bm k}-\varphi_{\bm p}$.
This equation is solved by expansion of the Cooperon in series of Fourier harmonics of the angle $\theta$~\cite{ILP,Knap,WL_ftp}.
The resonant at $q=0$ solution reads
\begin{equation}
C(\bm p', \bm p, \bm q)  = {\rm e}^{-i\theta}\mathcal C(\bm q),
\end{equation}
where the Cooperon $\mathcal C(\bm q)$ satisfies the equation
\begin{multline}
\mathcal C(\bm q) = \tau_0 
+ \qty[1-{2\pi g\tau_0\over \hbar}\expval{{1\over 4}{\rm Tr}\qty(U_{\bm p', \bm k}^\dag U_{\bm p', \bm k})-U_{\bm p', \bm k} \otimes U_{-\bm p', -\bm k} }_{\varphi_{\bm p'},\varphi_{\bm k}}]\mathcal C(\bm q)
\\- \tau_0\qty{
{\tau_{\rm tr}\over 2}
\qty( v\bm q + {2\gamma\over \hbar}[\bm S \times \hat{\bm z}])^2
+ {\tau_0\over 2}\qty({2\mu k^2\over \hbar})^2
J_z^2 + \Gamma_\phi} \mathcal C(\bm q).
\end{multline}
Here we took into account that the total departure rate is given by
\begin{equation}
\tau_0^{-1} = {2\pi g\over \hbar}\qty[{u^2 \over 2} + \expval{{1\over 4}{\rm Tr}\qty(U_{\bm p', \bm k}^\dag U_{\bm p', \bm k})}_{\varphi_{\bm p'},\varphi_{\bm k}}],
\end{equation}
and that the relaxation times of the first and third 
angular harmonics are equal to the transport time $\tau_{\rm tr}$ and the departure time $\tau_0$, respectively (they are related by $\tau_0 \approx \tau_{\rm tr}/2$).

Introducing the dephasing operator
\begin{equation}
\Gamma = \Gamma_wJ_z^2 + {2\pi g\over \hbar}\expval{{1\over 4}{\rm Tr}\qty(U_{\bm p', \bm p}^\dag U_{\bm p', \bm p})-U_{\bm p', \bm p} \otimes U_{-\bm p', -\bm p} }_{\varphi_{\bm p'},\varphi_{\bm p}},
\end{equation}
where $\Gamma_w=2(\mu \epsilon_{\rm F}^2/\hbar v^2)^2\tau_0$,
we can rewrite the Cooperon equation in the form
\begin{equation}
\qty[D\qty(\bm q-{2e\over \hbar}\bm A_{\rm SO})^2+\Gamma+\Gamma_\phi]\mathcal C(\bm q)=1,
\end{equation}
with the spin-orbit vector potential ${\bm A_{\rm SO} = (\gamma/ev) \hat{\bm z}\times \bm S }$.
It can be also rewritten  in the form of Eq.~(5) of the main text:
\begin{equation}
\label{SM_C_Eq}
\qty{Dq^2+\Gamma +\Gamma_\phi + \Gamma_{\rm R} \qty(\bm S^2- S_z^2) +\sqrt{2 \Gamma_{\rm R}\tau_{\rm tr}} [\bm S \times \bm q]_z v}\mathcal C(\bm q)=1,
\end{equation}
where $\Gamma_{\rm R}=2(\gamma/\hbar)^2\tau_{\rm tr}$.
It is important that 
the bilinear in $\bm S$ and $\bm q$ term is present in the Cooperon equation.

For the scattering amplitude given by Eq.~(4) of the main text we have
\begin{equation}
U_{\bm p', \bm p} = \cos({\theta/ 2}) \sum_{l,j=x,y,z}\beta_{lj}\Pi_l s_j
+\sum_{i,i'=x,y,z} \varkappa_i (\alpha_{ii'}s_{i'} + u_{ii'}\Pi_{i'}) + u{\varkappa_z \over \epsilon_{\rm F}} (\lambda s_z+\Delta_s \Pi_z).
\end{equation}
It is important that, by contrast to other terms, the  term $\propto \beta_{lj}$ in the scattering amplitude does not change sign at an interchange $\bm p', \bm p \to -\bm p', -\bm p$. 
Assuming a ``diagonal'' disorder~\cite{MF}: $\expval{u_{ii'}u_{nn'}}=u_{ii'}^2\delta_{in}\delta_{i'n'}$, $\expval{\alpha_{ii'}\alpha_{nn'}}=\alpha_{ii'}^2\delta_{in}\delta_{i'n'}$, $\expval{\beta_{lj}\beta_{l'j'}}=\beta_{lj}^2\delta_{ll'}\delta_{jj'}$ and isotropy in the $(xy)$ plane, and
using $\expval{\cos^2({\theta/ 2})}_{\varphi_{\bm p'},\varphi_{\bm p}}=\expval{\varkappa_i^2}_{\varphi_{\bm p'},\varphi_{\bm p}}=1/2$, we obtain
\begin{equation}
\Gamma = 
\Gamma_wJ_z^2 + {2\pi g\over \hbar}  \sum_{l,j=x,y,z}\qty[ \alpha_{lj}^2S_j^2 + u_{lj}^2 J_j^2+\beta_{lj}^2(S_j^2+ J_l^2-2S_j^2J_l^2)] + {2\pi g\over \hbar} \qty({u \over \epsilon_{\rm F}})^2 \qty(\lambda^2 S_z^2+\Delta_s^2 J_z^2) .
\end{equation}

For the valley singlet channel we have $J_{x,y,z}=0$, while for valley triplet channels we have either $J_z^2=0$, $J_x^2+J_y^2=2$ ($t_0$ channel) or $J_z^2=1$, $J_x^2+J_y^2=1$ ($t_1$ channel). Similarly, we have spin interference channels $s$, $t_0$ and $t_1$ with analogous expectation values of $S_{x,y,z}^2$ operators.
In total we have nine dephasing rates $\Gamma^{l}_{j}$ for channels with $l,j=s,t_0,t_1$. They are presented in Table~I of the main text.

\section{Magnetoconductivity calculation}

In magnetic field, it is convenient to search the Cooperon in the basis of Landau levels of a charge $2e$. Then the conductivity correction~\eqref{SM_cond} is rewritten as follows
\begin{equation}
\label{SM_cond_1}
\sigma=- \sigma_0 \sum_{n=0}^{N_0} \sum_{l=t_1,t_0,s} c_l 
\qty[ 
\mathcal C^l_s(n) - \sum_{m=1,0,-1} \mathcal C^l_{mm}(n) - \mathcal C^l_0].
\end{equation}
Here $n$ enumerates the Landau levels (for the triplet channel the Landau-level numbers are equal to $n+1$), $\mathcal C^l_0$ is a triplet contribution of the lowest Landau level, $N_0=\mathcal B_{\rm tr}/B \gg 1$ is the cutoff, and we use a representation of a total angular momentum of interferring particles with projections $m$ in the triplet channel.

The solution of the Cooperon Eq.~\eqref{SM_C_Eq} for the singlet channel where $\bm S=0$ has a simple form:
\begin{equation}
\mathcal C^l_s(n) = {1\over n+1/2+ b_\phi+b_s^l}, \qquad  b^l_j={\mathcal B^l_j\over B}, \qquad  b_\phi={\mathcal B_\phi\over B}.
\end{equation}
For triplet Cooperons we have systems of equations with off-diagonal elements caused by the $\bm S$-linear terms in Eq.~\eqref{SM_C_Eq}. Therefore the triplet Cooperon $\mathcal C^l_{mm'}(n)$  is given by elements of inverse matrices
\begin{equation}
\mathcal C^l(n \geq 1) = \qty[\mathcal L^l(n)]^{-1}, \qquad \mathcal C^l(n=0) = \qty[\mathcal L^l_0]^{-1}, 
\end{equation}
where~\cite{ILP,Knap,Punnoose}
\begin{equation}
\label{triplet_matrix}
\mathcal L^l(n)=
\begin{pmatrix}
\epsilon_{n-1}+b_{\rm R} - b_-& i\sqrt{2 b_{\rm R} n} &0
\\ -i\sqrt{2 b_{\rm R} n} & \epsilon_n+2b_{\rm R} + 2 b_-&i\sqrt{2 b_{\rm R} (n+1)}
\\ 0 & -i\sqrt{2 b_{\rm R} (n+1)} & \epsilon_{n+1}+b_{\rm R} - b_-
\end{pmatrix},
\quad
\mathcal L^l_0=
\begin{pmatrix}
\epsilon_0+2b_{\rm R} + 2 b_-&i\sqrt{2 b_{\rm R}}
\\-i\sqrt{2 b_{\rm R}} & \epsilon_{1}+b_{\rm R}- b_- 
\end{pmatrix}
\end{equation}
with 
\begin{equation}
\epsilon_n = n+1/2 + b_\phi+\bar{b}, \qquad \bar{b} = {b_{0}+2b_{1}\over 3}, \qquad b_-={b_{0}-b_{1}\over 3},
\qquad b_{1,0}={\mathcal B^l_{t_1,t_0} \over B}.
\end{equation}
The triplet lowest Landau level Cooperon is $\mathcal C^l_0=1/( \epsilon_0+b_{\rm R}-b_-)$.

Calculating the trace of the matrix $\qty[\mathcal L^l(n)]^{-1}$ we obtain
\begin{equation}
{\sigma(B) \over \sigma_0} = -\sum_{l=t_1,t_0,s}c_l  \qty[\sum_{n=0}^{N_0} \qty( {1\over n+1/2+ b_\phi+b_s^l} - {3\epsilon_n^2 + 4b_{\rm R}\epsilon_n + A \over\epsilon_n^3 + C\epsilon_n +2G})+{1\over \epsilon_0+b_{\rm R}-b_- -1} - {1\over \epsilon_0+b_{\rm R}-b_-}],
\end{equation}
where we 
added and subtracted the term ${\rm Tr} \qty{\qty[ \mathcal L^l(0)]^{-1}}$ to Eq.~\eqref{SM_cond_1} and 
used the relation ${\rm Tr} \qty{\qty[ \mathcal L^l(0)]^{-1}}-{\rm Tr}\qty[\qty(\mathcal L^l_0)^{-1}]={1/(\epsilon_0+b_{\rm R}-b_- -1)}$. 
The coefficients $A$, $C$, $G$ dependent on $b_{\rm R}$, $b_\phi$, $b_0$ and $b_1$ are given in the main text.
This sum can be evaluated owing to the expansion~\cite{Punnoose}
\begin{equation}
{3\epsilon_n^2 + 4b_{\rm R}\epsilon_n + A \over \epsilon_n^3 + C\epsilon_n +2G}
= \sum_{m=0,\pm}{u_m \over \epsilon_n - v_m},
\end{equation}
where the coefficients $u_{0,\pm}$ and $v_{0,\pm}$ are expressed via $A$, $C$, $G$ and $b_{\rm R}$,  see the main text.
Calculating the magnetoconductivity $\Delta\sigma=\sigma(B)-\sigma(0)$ we obtain
\begin{equation}
{\Delta\sigma \over \sigma_0} 
= -\sum_{l=t_1,t_0,s} c_l \biggl[ \mathcal F_t\qty({b_\phi},{b_{\rm R}},{b_{1}},{b_{0}}) 
\\ - F\qty({b_\phi +  b_{s}^{l}}) \biggr],
\end{equation}
where $F(x)=\psi(1/2+1/x)+\ln{x}$ is the HLN function  with the digamma function $\psi(y)$, and
\begin{equation}
\mathcal F_t\qty(b_\phi,b_{\rm R},b_1,b_0) = 
\sum_{m=0,\pm}  \qty[u_m \psi\qty(1/2 + b_\phi  + \bar{b} - v_m)  - u_m^{(0)}\ln{\qty(b_\phi  + \bar{b} - v_m^{(0)})}]
+{1\over (b_\phi+b_{\rm R} + b_{1})^2-1/4}.
\end{equation}
The coefficients $v_m^{(0)}$ and $u_m^{(0)}$ are the zero-field asymptotes of $v_m$ and $u_m$ calculated by the zero-field asymptotes of $A$, $C$ and $G$ given by $A^{(0)}=A+1$, $C^{(0)}=C+1$, $G^{(0)}=G+b_-$.
Note that at $b_- =0$, 
the term $\propto u_m^{(0)}$ 
can be expressed via logarithms and inverse trigonometric functions of real arguments,
see Refs.~\cite{wl_tilted} and~\cite{Punnoose}.

\end{document}